\documentclass[sigconf, nonacm]{acmart}

\usepackage{bbm}
\usepackage{appendix}

\usepackage{graphicx}
\usepackage{caption}
\usepackage{subcaption}

\usepackage[inline]{enumitem}
\usepackage{wrapfig}

\usepackage{footnote}

\usepackage{acronym}
\usepackage{mathtools}

\newlist{questions}{enumerate}{2}
\setlist[questions,1]{label=Q\arabic*.,ref=Q\arabic*}
\setlist[questions,2]{label=(\alph*),ref=\thequestionsi(\alph*)}

\makesavenoteenv{tabular}
\makesavenoteenv{table}

\acrodef{JS}{Jensen Shannon}

\allowdisplaybreaks
\copyrightyear{2022} 
\acmYear{2022} 
\setcopyright{acmlicensed}
\acmConference[RecSys '22]{Sixteenth ACM Conference on Recommender Systems}{September 18--23, 2022}{Seattle, WA, USA}
\acmBooktitle{Sixteenth ACM Conference on Recommender Systems (RecSys '22), September 18--23, 2022, Seattle, WA, USA}
\acmPrice{15.00}
\acmDOI{10.1145/3523227.3546780}
\acmISBN{978-1-4503-9278-5/22/09}

\begin{document}

\title[RADio -- Rank-Aware Divergence Metrics to Measure Normative Diversity in News]{RADio -- Rank-Aware Divergence Metrics to Measure Normative Diversity in News Recommendations}

\author{Sanne Vrijenhoek}
\affiliation{
    \institution{Institute for Information Law, University of Amsterdam}
    \city{Amsterdam}
    \country{The Netherlands}
}
\email{s.vrijenhoek@uva.nl}
\authornote{These authors contributed equally.}

\author{Gabriel Bénédict}
\affiliation{
    \institution{University of Amsterdam, RTL Nederland B.V.}
    \city{Amsterdam}
    \country{The Netherlands}
}
\email{gabriel.benedict@rtl.nl}
\authornotemark[1]

\author{Mateo Gutierrez Granada}
\affiliation{
    \institution{RTL Nederland B.V.}
    \city{Hilversum}
    \country{The Netherlands}
}
\email{mateo.gutierrez.granada@rtl.nl}

\author{Daan Odijk}
\affiliation{
    \institution{RTL Nederland B.V.}
    \city{Hilversum}
    \country{The Netherlands}
}
\email{daan.odijk@rtl.nl}

\author{Maarten de Rijke}
\orcid{0000-0002-1086-0202}
\affiliation{
    \institution{University of Amsterdam}
    \city{Amsterdam}
    \country{The Netherlands}
}
\email{m.derijke@uva.nl}

\renewcommand{\shortauthors}{Vrijenhoek et al.}

\begin{abstract}

In traditional recommender system literature, diversity is often seen as the opposite of similarity, and typically defined as the distance between identified topics, categories or word models. However, this is not expressive of the social science's interpretation of diversity, which accounts for a news organization's norms and values and which we here refer to as \emph{normative} diversity. We introduce RADio, a versatile metrics framework to evaluate recommendations according to these normative goals. 

RADio introduces a rank-aware \ac{JS} divergence. This combination accounts for (i) a user's decreasing propensity to observe items further down a list and (ii) full distributional shifts as opposed to point estimates. 
We evaluate RADio's ability to reflect five normative concepts in news recommendations on the Microsoft News Dataset and six (neural) recommendation algorithms, with the help of our metadata enrichment pipeline. We find that RADio provides insightful estimates that can potentially be used to inform news recommender system design. %

\end{abstract}

\keywords{News recommendation, Diversity, Divergence, Normative framework}

\maketitle

\acresetall

\section{Introduction}
For centuries, the interplay between journalists and news editors has shaped how news items are created and how they are shown to their readers~\cite{newsrooms}. With the digitization of society, much has changed: while before, people would typically limit themselves to reading one type of newspaper, they now have a wealth of information available to them at the click of a button~\cite{usersChoosePopularity} -- more than anyone could possibly be expected to read or make sense of.
News recommender systems can filter the enormous amount of information available to just those news items that are in some way interesting or relevant to their users~\cite{moller2018not,bodo2019}. The use of news recommender systems is widespread, not just for \emph{personalized} news recommendations, but also to automatically populate the front page of a news website~\cite{doi:10.1080/1461670X.2022.2034522}, or present the reader of a particular news article with other articles about the same topic, but from a different perspective~\cite{mulder2021operationalizing}. 
The use of news recommender systems has a wide range of benefits. They can increase engagement~\cite{nic2018reuters} and help raise informed citizens~\cite{eskens2017}. A news recommender system may broaden the horizons of their users by presenting diverse recommendations, including items different from what they are used to or expect seeing. They could even foster tolerance and understanding~\cite{ferre2018,stromback2005}, and counter so-called filter bubbles or echo chambers ~\cite{pariser2011filter,moller2018not}.

\begin{figure*}[t]
\centering
\begin{subfigure}{\linewidth}
  \centering
  \includegraphics[width=1\linewidth]{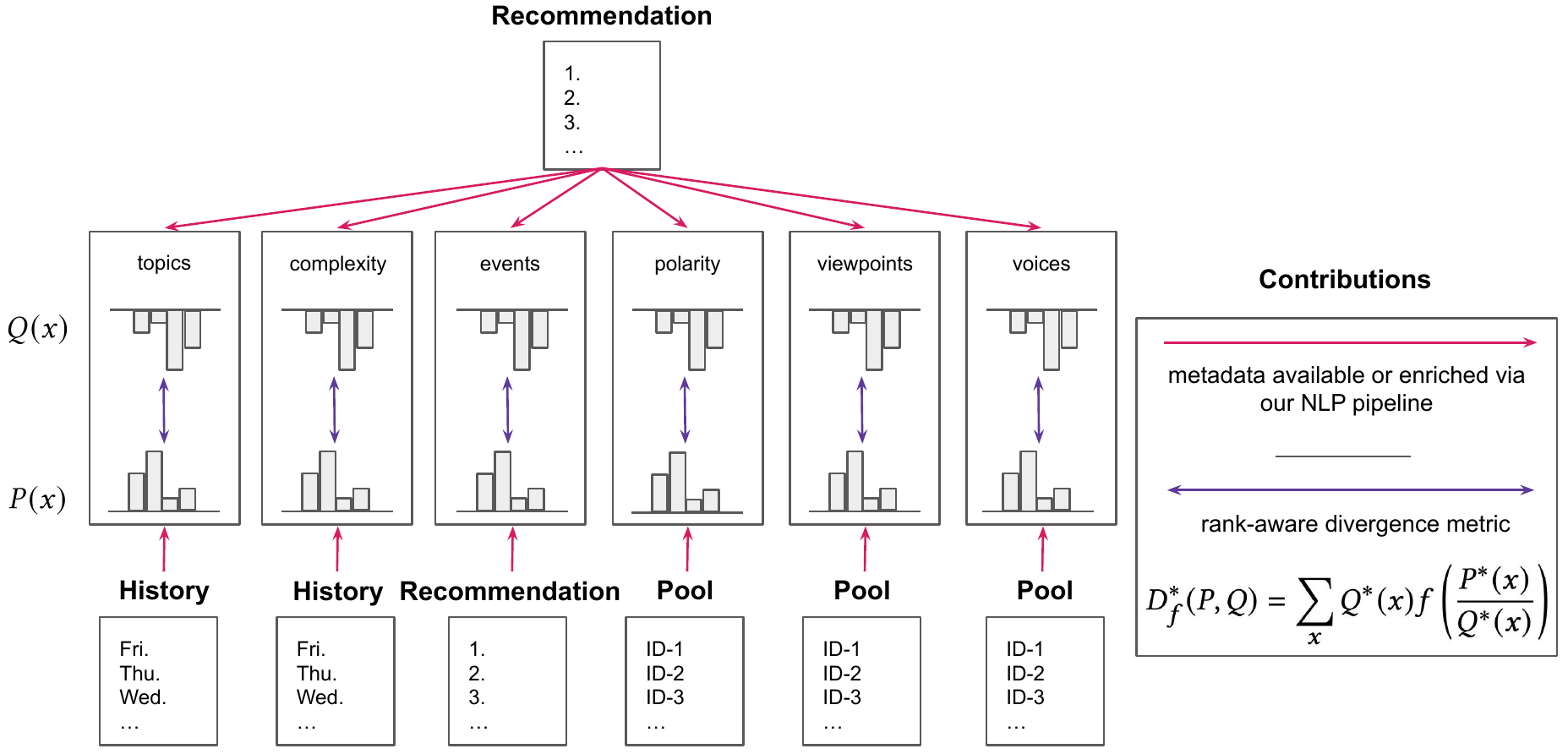}
  \label{fig:pvsq1}
\end{subfigure}%
\vspace*{-\baselineskip}
\caption{Comparing discrete diversity distributions in the context of news recommendations. First, metadata is collected in the news dataset or retrieved via our NLP pipeline (red). Discrete distributions of that metadata are then compared via a rank-aware divergence metric (purple). Recommendation set $Q$ and the context articles $P$ are compared with rank-aware f-Divergence.}
\label{fig:dist}
\end{figure*}

To realize the potential benefits of news recommender systems, much attention has been given to generating recommendations that reflect the user's interests and preferences~\cite{karimi2018news}. However, with news recommenders taking over the role of human editors in news selection, they are becoming gatekeepers in what news is shown to audiences and have thus a democratic role to play in society. As such, their evaluation has different requirements than those of other types of recommender systems~\cite{beam2014automating,wallace2018modelling,welbers2018gatekeeper,bastian2019safeguarding}. Recent controversies have shown that merely optimizing for click-through rates and engagement may promote sensationalist content~\cite{tenenboim2015prompts}, and is particularly conducive to the spread of misinformation.\footnote{See, for example, the alleged role Facebook played in the storming of the Capitol: \url{https://www.washingtonpost.com/technology/2021/10/22/jan-6-capitol-riot-facebook/}} 
This observation is not limited to the academic literature -- an increasing number of media organizations, both public service and commercial, have
acknowledged the difficulties in translating their editorial norms into concrete metrics that can inform recommender system design~\cite{grunchallenges2021,boididou2021building}.
News recommender systems exist in a complex space consisting of many different areas and disciplines, each with their own goals and challenges; think of balancing diversity and accuracy~\cite{10.1145/3460231.3474234}, nudging~\cite{mattis2021nudging} or even identifying user preferences~\cite{10.1145/3209978.3210007,bernheim2021theory} and biases~\cite{10.1145/3460231.3474244}. 
In this paper, we focus on the process of translating normative theory (i.e., what it means for a recommendation to be diverse) into metrics that are usable and understandable for both technical and editorial purposes. We build on the work of \citet{helberger2019democratic_rs}, who provides a theoretical foundation for conceptualizing diversity, and of \citet{vrijenhoek2021recommenders}, who propose a new set of metrics (DART) that reflect this theory. The DART metrics represent a first step towards a normative interpretation of diversity in news recommendations. We identify a number of possible shortcomings in these metrics: there could be more consideration for the theory of metrics and distance functions, generalizability to other normative concepts, unification under one framework, and rank-awareness. 
In this paper, we focus on the mathematical aspects of a rank-aware metric, versatile to different normative concepts and as such addressing these shortcomings. We refer to our framework as 
the \emph{Rank-Aware Divergence metrIcs to measure nOrmative diversity} (RADio).

Our contribution consists of a diversity metric that is (i) versatile to any normative concept and expressed as the divergence between two (discrete) distributions; (ii) rank-aware, taking into account the position of an item in a recommendation set; and %
(iii) mathematically grounded in distributional divergence statistics. We demonstrate the effectiveness of this formulation of the metrics by defining a natural language processing (NLP) metadata enrichment pipeline (e.g., sentiment analysis, named entity recognition) and running it against the MIND dataset~\cite{wu2020mind}. Figure \ref{fig:dist} illustrates the operationalization. The pipeline and the code produced for metadata enrichment and metric computation are available online.\footnote{\url{https://github.com/svrijenhoek/RADio}} The goal of RADio is not to serve as thresholds or strict guidelines for ``diverse recommendations,'' but to provide developers of recommender systems with the tools to evaluate their systems on normative principles.

\label{sec:introduction}

\section{Related Work}
\label{sec:background}

We first highlight recent work on the formal mathematical work on diversity in news recommendation, before citing related work on the normative aspect of diversity. Finally we describe the gap that exists between descriptive and normative diversity.\footnote{This dichotomy is oftentimes referred to as normative (\emph{what ought to be}) and positive (\emph{what is}) statements \cite{hume} but can easily be confused with concepts such as positive / negative examples in Machine Learning. We thus opt for the more explicit normative / descriptive duo.}

\subsection{Descriptive (General-Purpose) Diversity}
\label{sec:posDiv}

\begin{table*}[t!]
\caption{Overview of the different models and expected value ranges for each metric. It should be noted that a high score should be interpreted as high \emph{divergence}; As such, a high score does not necessarily mean a better score.}
\label{tab:table_overview}
\resizebox{\textwidth}{!}{%
\begin{tabular}{l cccccc}
\toprule
\textbf{} & \textbf{\begin{tabular}[c]{@{}c@{}}Calibration\\ (topic)\end{tabular}} & \textbf{\begin{tabular}[c]{@{}c@{}}Calibration\\ (complexity)\end{tabular}} & \textbf{Fragmentation} & \textbf{Activation} & \textbf{Representation} & \textbf{Alternative voices} 
\\
\midrule
\textbf{Liberal} & Low & Low & High & -- & -- & -- 
\\
\textbf{Participatory} & High & Low & Low & Medium & Reflective & Medium \\ 
\textbf{Deliberative} & -- & -- & Low & Low & Equal & -- 
\\
\textbf{Critical} & -- & -- & -- & High & Inverse & High \\
\bottomrule
\end{tabular}
} %
\end{table*}

Diversity is a central concept in Information Retrieval literature~\cite{Clarke2008IRDiv, Sakai2019GoodDiversity}, albeit with a different interpretation than the normative diversity described in the previous section. During the development of news recommender systems, there is currently a large focus on the predictive power of an algorithm. However, this may unduly promote content similar to what a user has interacted with before, and lock them in loops of ``more of the same.''%
{} To tackle this, ``diversity'' is introduced, which is typically defined as the ``opposite of similarity'' \cite{Bradley2001ImprovingRD}. Its goal is to prevent users from being shown the same type of items in their recommendations list and is often expressed as intra-list-diversity (ILD) \cite{Bradley2001ImprovingRD, DiNoia2014UserDiversity, Ekstrand2014UserPerceptionDiv, JUGOVAC2017321, Du2021diversitySuitable, vargas2011rank,castells2015novelty,lu2020beyond}: mean pairwise dissimilarity between recommended item lists. ILD requires the specification of a distance function between lists, and thus leaves it up to interpretation as to what it means for two lists to be distant. In theory, it could still be interpreted with a metric that accounts for the presence of different sources or viewpoints~\cite{pmlr-v81-ekstrand18b}. However, in practice, diversity is most often implemented as a descriptive distance metric such as cosine similarity between two bag-of-words models or word embeddings~\cite{lu2020beyond,kunaver2017}.

Other popular ``beyond-accuracy'' metrics related to diversity are novelty (how different is this item from what the user has seen in the past), serendipity (is the user positively surprised by this item), and coverage (what percentage of articles are recommended to at least one user). %
These metrics can be taken into account at different points in the machine learning pipeline~\cite{kunaver2017,Wu2019RecentAI}. One can optimize for these descriptive notions of diversity (i) before training, by clustering users based on their profile diversity with JS divergence \cite{Eskandanian2017ClusteringDiverse},
(ii) directly at training time (e.g., for learning-to-rank \cite{Borodin2012DiverseRanking, vargas2011rank, castells2015novelty}, collaborative filtering \cite{qin2013CFDiversity}, graphs \cite{Gan2020GraphsOpt, Puthiya2016diversereco} or bandits \cite{Ding2021BanditOpt, Xie2021NetworkOpt}), (iii) by re-ranking a recommendation set and balance diversity vs.\ relevance~\cite{Chen2018NIPSReranking} or popularity vs.\ relevance~\cite{chakraborty2019equalityVoice}, 
and (iv) by defining a post-recommendation metric to measure diversity for each recommendation set or at user-level (e.g., the generalist-specialist score \cite{waller2019GSscore, Anderson2020AlgorithmicEO}). With any of these four methods, a trade-off must be made between the relevance of a recommendation issued to users and the level of descriptive diversity, though there have also been studies indicating that increasing diversity does not necessarily need to negatively affect relevance~\cite{lu2020beyond}. Nevertheless, this encouraged recent efforts in training neural-based recommenders that explicitly make a trade-off between accuracy and diversity~\cite{raza2021DNNDivAcc}. Also recently, there have been studies that differentiate between diversity needs of users~\cite{wu2018personalizing}. %

\subsection{Normative Diversity}
\label{sec:normDiv}
Diversity is extensively discussed as a normative concept in literature, and has a role in many different areas of science~\cite{steel2018multiple,loecherbach2020unified}, spanning from ecological diversity to diversity as a proxy for fairness in machine learning systems~\citep{mitchell2020diversity}. While these interpretations of diversity are often related, they do not fully cover the nuances of a diverse \emph{news} recommender system, the work on which stems from democratic theory and the role of media in society.
Following \citet{helberger2019democratic_rs}, we define a normatively diverse news recommendation as one that succeeds in informing the user and supports them in fulfilling their role in democratic society. 
Out of the many theoretical models that exist in literature, \citet{helberger2019democratic_rs} describes four different models from the normative framework of democracy, each with a different view on what it means to properly inform citizens: the \textbf{Liberal} model, which aims to enable personal development and autonomy, the \textbf{Participatory} model, which aims to enable users to fulfill their role as active citizens in a democratic society, the \textbf{Deliberative} model, which aims to foster discussion and debate by equally presenting different viewpoints and opinions in a rational and neutral way, and the \textbf{Critical} model, which aims to challenge the status quo and to inspire the readers to take action against existing injustices in society.

\noindent%
For more details regarding the different models, and what a recommender system following each of these models would look like, we refer to~\citet{helberger2019democratic_rs}. Which model is followed is a decision that needs to be made by the media organization itself, and should be in line with their norms and values. 

Based on these models, the DART
metrics \cite{vrijenhoek2021recommenders} take a first step towards normative diversity for recommender systems and reflect the nuances of the different democratic models described above: \textbf{Calibration}, \textbf{Fragmentation}, \textbf{Activation}, \textbf{Representation} and \textbf{Alternative Voices}. Table \ref{tab:table_overview} provides an overview of the DART metrics and their expected value ranges for the different models, and will be further elaborated later in the paper.%

\subsection{The Gap Between Normative and Descriptive Diversity}
The descriptive diversity metrics described in Section~\ref{sec:posDiv} are general-purpose and meant to be applicable in all domains of recommendation. However, in their simplicity a large gap can be observed between this interpretation of diversity and the social sciences' perspective on media diversity that is detailed in Section~\ref{sec:normDiv}. In their comprehensive work on the implementation of media diversity across different domains, \citet{loecherbach2020unified} note that there is ``little to no overlap between concepts and operationalizations (of diversity) used in the different fields interested in media diversity.'' 
As such, a recommendation that would score high on diversity according to traditional information retrieval-based metrics \cite{Clarke2008IRDiv, Sakai2019GoodDiversity}, may not be considered to be diverse according to the criteria maintained by newsroom editors. Both \citet{loecherbach2020unified} and \citet{bernstein2020diversity} call for truly interdisciplinary research in bridging this gap, where \citet{bernstein2020diversity} argue for close collaboration between academia and industry and the foundation of joint labs. 
This work is a step in that direction, as we provide a versatile and mathematically grounded rank-aware metric that can be used by practitioners to monitor their normative goals.

\section{Operationalizing Normative Diversity for News Recommendation}
\label{sec:method}
With our RADio framework, we further refine the DART metrics that were defined by \citet{vrijenhoek2021recommenders} in order to resolve a number of the shortcomings of the metrics' initial formalizations. In their current form, each of the metrics has different value ranges; for example, \emph{Activation} has a value range $[-1,1]$, where a higher score indicates a higher degree of activating content, and \emph{Calibration} has a range of $[0,\infty]$, where a lower score indicates a better Calibration. These different value ranges reduce the interpretability of the metrics, making them harder to explain and as such less likely to be adopted by news editors.
Furthermore, the proposed metrics do not take the position of an article in a recommendation into account. News recommendations are ranked lists of articles that are typically presented to users in such a way that the likelihood of a recommended article to be considered by the user decreases further down the ranking. As such, in the evaluation of the diversity of the recommender system we should also account for the position of an article in the recommendation ranking, rather than considering the set as a whole (e.g. ILD). 

Thus, the two major challenges that we seek to address are that (i) scores should be comparable between the metrics and across recommendation systems, and (ii) scoring of both unranked and ranked sets of recommendations should be possible. In this section, we first detail these requirements (Section~\ref{sec:metrics_requirements}), then describe how we reformulate the metrics to each use the same divergence-based approach (Section~\ref{sec:metrics_divergence}). We then add the rank-aware aspect to the metrics (Section~\ref{sec:metrics_ranking}), before applying them to the five concrete DART metrics (Section~\ref{sec:metricsinpractice}).

\subsection{Requirements}
\label{sec:metrics_requirements}

We first enunciate the classical definition of a distance metric, before specifying three desirable metric criteria for news recommendations. Take a set $X$ of random variables and $x, y, z \in X$, then a metric $D$ is a proper distance measure if ${D(x, y)=0 \Leftrightarrow x=y}$, ${D(x, y) = D(y, x)}$ and ${D(x, y) \leq D(x, z)+D(z, y)}$. These are respectively the axioms of \emph{identity}, \emph{symmetry} and \emph{triangle inequality}, that express intuitions about concepts of distance \cite{Searcoid2007distanceMetrics}. 

We add that our distance measure should (i) be bounded by $[0;1]$, for comparisons of different recommendation algorithms (ii) be unified, so as to fairly consider different diversity aspects (as opposed to e.g. using weighted averages or maxima in \citep{Dhamala2021fairMetrics}) and (iii) allow for discrete rank-based distribution sets, to fit the ranked recommendation setting.

\subsection{f-Divergence}
\label{sec:metrics_divergence}

We model the task of measuring diversity as a comparison between probability distributions: \emph{the difference in distribution between the issued recommendations} ($Q$) \emph{and its context} ($P$). Each diversity metric prescribes its own $Q$ and $P$. The elements in the distribution $Q$ can be recommendation items (cf.\ Calibrated Recommendations \cite{steck2018}), but can also be higher-level concepts, such as distributions of topics and viewpoints. The context $P$ may refer to either the overall supply of available items, the user profile, such as the reading history or explicitly stated preferences, or the recommendations that were issued to other users (see Figure \ref{fig:dist}). Intuitively, when $P$ is linked to the same user as $Q$, we measure within user diversity (e.g., towards preventing getting locked in ``filter bubbles''). When $P$ is linked to another user than $Q$, we measure diversity across users (e.g., monitoring diversity of viewpoints represented across personalized homepages). In the following, we formalize the role of $P$ and $Q$ in two different metric settings, starting with the simple and common KL divergence metric, before presenting its refinement (Jensen-Shannon divergence) as our preferred metric.

\subsubsection{Kullback-Leibler Divergence}
\label{sec:KLDefinition}
The concept of relative entropy or KL (Kullback–Leibler) divergence ~\cite{Kullback1951KL-divergence} between two probability mass functions $P$ and $Q$ (here, a recommendation and its context) is defined as:
\begin{equation}
D_{\mathrm{KL}}(P, Q)=-\sum_{x \in \mathcal{X}} P(x) \log_{2}Q(x) + \sum_{x \in \mathcal{X}} P(x) \log_{2}P(x).
\end{equation}
Often also expressed as $D_{\mathrm{KL}}(P, Q) = H(P,Q) - H(P)$, with $H(P,Q)$ the cross entropy of $P$ and $Q$, and $H(P)$ the entropy of P. Both cross entropy and KL divergence can be thought of as measurements of how far the probability distribution $Q$ is from the reference probability distribution $P$. When $P=Q$, $D_{\mathrm{KL}}(P, Q) = D_{\mathrm{KL}}(P, P) = 0$, that identity property is not guaranteed by cross entropy alone. This is the main reason to prefer KL divergence over cross entropy. Though KL Divergence satisfies the \emph{identity} requirement, the \emph{symmetry} and \emph{triangle inequality} are not fulfilled. This can be resolved by further refining KL Divergence.

\subsubsection{Jensen–Shannon Divergence}
\label{sec:SJDefinition}

A succession of steps from KL divergence lead to Jensen-Shannon (JS) divergence. KL divergence was first turned symmetric~\cite{jeffreys1946invariant} and then upper bounded~\cite{lin1991divergence}, to lead to

\begin{equation}
\begin{multlined}
D_{\mathrm{JS}}(P, Q) =
 -\sum_{x \in \mathcal{X}} \frac{P(x)+Q(x)}{2} \log_{2}\left(\frac{P(x)+Q(x)}{2}\right) \\
+ \frac{1}{2}\sum_{x \in \mathcal{X}} P(x) \log_{2}P(x) + \frac{1}{2}\sum_{x \in \mathcal{X}} Q(x) \log_{2}Q(x)
\end{multlined}
\end{equation}
\noindent%
When the base $2$ logarithm is used, the JS divergence bounds are $0 \leq D_{\mathrm{JS}}(P, Q) \leq 1$. Additionally, \citet{endres2003new} show that $\sqrt{D_{\mathrm{JS}}}$ is a proper distance which fulfills the identity, symmetry and the triangle inequality properties. When we refer to $D_{\mathrm{JS}}$ or JS divergence below, we therefore implicitly refer to the square root of the JS formulation with log base 2.

\citet{Liese2006fDiv} defined \emph{f-Divergence} [$D_f$]: a generic formulation of several divergence metrics. Among them are the JS and KL divergences.\footnote{f-Divergence accommodates for other divergence metrics which are out of scope of this research \cite{Liese2006fDiv}.} Further along the text, we use $D_f$ as a shorthand notation for KL and JS divergences. $D_f$ in discrete form is
\begin{equation}
D_f (P,Q) = \sum_x Q(x) f \left( \frac{P(x)}{Q(x)} \right),
\end{equation}
where $f_{\mathrm{KL}}(t) = t \log t $ and $f_{\mathrm{JS}}(t) = \frac{1}{2}\left[(t+1) \log \left(\frac{2}{t+1}\right)+t \log t\right]$.
To avoid misspecified metrics~\citep{steck2018}, we write $\overline{P}$ and $\overline{Q}$:
\begin{equation}
\overline{Q}(x) = (1-\alpha) Q(x) + \alpha P(x) \quad \overline{P}(x) = (1-\alpha) P(x) + \alpha Q(x),
\end{equation}
where $\alpha$ is a small number close to zero. $\overline{P}$ prevents artificially setting $D_f$ to zero when a category  (e.g., a news topic) is represented in $Q$ and not in $P$. In the opposite case (when a category is represented in $P$ and not in $Q$), $\overline{Q}$ avoids zero divisions. In order for the entire probabilistic distributions $\overline{P}$ and $\overline{Q}$ to remain proper statistical distributions, we normalize them to ensure $\sum_x \overline{P}(x) = \sum_x \overline{Q}(x) = 1$. To avoid notation congestion, $P$ and $Q$ will implicitly refer to $\overline{P}$ and $\overline{Q}$, in the following sections.

\subsection{Rank-Aware f-Divergence Metrics}
\label{sec:metrics_ranking}

Our ranked recommendation setting (characteristic (iii) above) motivates a further reformulation of our f-Divergence metric. It is well entrenched in Learning To Rank (LTR) literature \cite{Tax2015metaLTR, Yilmaz2010effectivenessLTR}, and by extension in conventional descriptive diversity metrics~\cite{castells2015novelty} that a user is a lot less likely to see items further down a recommended ranked list (i.e., diminishing inspection probabilities). Note that the ranking oftentimes reflects relevance to the user, but it is not always the case for news (e.g., editorial layout of a news homepage).

We extend our metrics with an optional discount factor for $P$ and $Q$ to weigh down the importance of results lower in the ranked recommendation list. The ranking relevancy metrics Mean Reciprocal Rank (MRR) and Normalized Discounted Cumulative Gain (NDCG) are popular rank-aware metrics for LTR \cite{Jarvelin2002NDCG, Chakrabarti2008MMRMAPNDCG}, in particular for news recommendation \cite{wu2020mind}. In line with the LTR literature, we first define the discrete probability distribution of a ranked recommendation set $Q^*$, given each item $i$ in the recommendation list $R$:
\begin{equation}
Q^*(x) = \frac{\sum_i w_{R_i} \mathbbm{1}_{i \in x}}{\sum_i w_{R_i}},
\end{equation}
where $w_{R_i}$, the weight of a rank for item $i$, can be different depending on the discount form. For MMR, $w_{R_i} = 1 / R_i$, for NDCG, $w_{R_i} = 1 / \log_2(R_i + 1)$. When $w_{R_i}$ = 1, $Q^*$ is not discounted (i.e., $Q^* = Q$).

In news recommendation, the \textit{sparsity bias} plays a predominant role: users will interact with a small fraction of a large item collection, such as scrollable news recommendation websites \cite{Kille2013plista}. We thus opt for weighing based on MRR rather than NDCG, because it applies a heavier discount along the ranking than NDCG. Note that the latter is said to be more suited for query-related rankings, where the user  has a particular information need related to a query and thus higher propensity to scroll down a page \cite{Chakrabarti2008MMRMAPNDCG}.

The context distribution $P$ is discounted in the same manner, when it is a ranked recommendation list. When $P$ is a user's reading history (see Figure~\ref{fig:dist}), the discount on $P$ increases with time: articles read recently are weighted higher than articles read longer ago. There are situations when rank-awareness is not applicable, for example  when $P$ is the entire pool of available articles.\footnote{There are several features along which such a pool of data could be ranked besides recency, such as the popularity during the last hour, day or week. As this is an editorial decision we remain agnostic as to the choice of that feature and refrain from ranking, though it remains possible in theory.} With rank-aware $Q^*$ and optionally rank-aware $P^*$, we formulate RADio, our rank-aware f-Divergence metric:
\begin{equation}
D^*_f (P,Q) = \sum_x Q^*(x) f \left( \frac{P^*(x)}{Q^*(x)} \right),
\end{equation}
$Q^*(x)$ and $P^*(x)$ accommodate for multiple situations: for example, $Q^*(c|R)$ is the rank-aware distribution of news categories $c$ over the recommendation set $R$. In the following, we specify $P^*(x|\cdot)$ and $Q^*(x|\cdot)$ in accordance to each normative concept of interest for our universal metric.

\label{sec:metrics_implementation}
\subsection{Normative Diversity metrics as Rank-Aware f-Divergences}
\label{sec:metricsinpractice}

In this section, we describe the RADio formalization of the general f-Divergence formulation above to the five DART metrics. We leave the exact implementation of the metrics in practice for a particular open news recommendation dataset to the next section. %
More formally, we define the following global parameters:
\begin{itemize}[leftmargin=*]
    \item $S$: The list of news articles the recommender system could make its selection from, also referred to as the ``supply.''
    \item $R$: The ranked list of articles in the recommendation set.
    \item $H$: The list of articles in a user's reading history, ranked by recency.
\end{itemize}

\noindent%
$R^{u}_{i} \in \{1, 2, 3, \ldots\}$ refers to the rank of an item $i$ in a ranked list of recommendations for user $u$. In this work, metrics are defined for a specific user at a certain point in time, therefore $R$ implicitly refers to $R^{u}$, unless stated otherwise. %
While this section contains some contextualization of the DART metrics \cite{vrijenhoek2021recommenders}, the original paper contains further normative justifications.

\paragraph{Calibration} (Equation \ref{eq:cal}) measures to what extent the recommendations are tailored to a user's preferences. The user's preferences are deduced from their reading history ($H$). Calibration can have two aspects: the divergence of the recommended articles' \emph{categories} and \emph{complexity}. The former is expected to be extracted from news metadata and thus categorical by nature, the latter is a binned (categorical) probabilistic measure extracted via a language model. As such, we compare $P^*(c|H)$, the rank-aware distribution of categories or complexity score bins $c$ over the users' reading history, and $Q^*(c|R)$ the same in the recommendations issued to the user.

\paragraph{Fragmentation} (Equation \ref{eq:frag}) reflects to what extent we can speak of a common public sphere, or whether the users exist in their own bubble. We measure Fragmentation as the divergence between every pair of users' recommendations. Here we consider $P^*(e|R^{u})$ as the rank-aware distribution of news events $e$ over the recommendations $R$ for user $u$, and $Q^*(e|R^{v})$ the same but for user $v$. KL Divergence is asymmetric (see Section \ref{sec:KLDefinition}), which means that its outcome differs depending on which user's recommendation is chosen as the target and which as the reference distribution. To avoid this, we compute the Fragmentation score as the average of KL Divergences with switched parameters. JS divergence is already symmetric and is thus implemented as for the other metrics.
In theory, Fragmentation requires a user's recommendation to be compared to those of all other users. This is not feasible with a sizeable dataset and the requirement of a reasonable compute time. Instead we opt to randomly sample user pairs. 

\paragraph{Activation} (Equation \ref{eq:act})
Most off-the-shelf sentiment analysis tools analyze a text, and return a value $(0,1]$ when the text expresses a positive emotion, a value $[-1,0)$ when the expressed sentiment is negative, and 0 if it is completely neutral. The more extreme the value, the stronger the expressed sentiment is. As proposed in \citep{vrijenhoek2021recommenders}, we use an article's absolute sentiment score as an approximation to determine the height of the emotion and therefore the level of Activation expressed in a single article. This then yields a continuous value between 0 and 1. $P(k|S)$ denotes the distribution of (binned) article Activation score $k$ within the pool of items that were available at that point ($S$). $Q^*(k|R)$ expresses the same, but for the binned Activation scores in the rank-aware recommendation distribution.

\begin{table*}[t]
\small
\caption{Overview of the implementation approach for different methods. Numbers in bold correspond to the corresponding steps in the metadata enrichment pipeline presented above.}
\label{tab:table_implementation}
\renewcommand{\arraystretch}{-10}
\begin{tabular}{llp{0.1\textwidth}p{0.45\textwidth}}
\toprule
\textbf{} & \textbf{Context} & \textbf{Type} & \textbf{Distribution of} 
\\ 
\midrule
\textbf{Calibration (topics)} & Reading history & Categorical & article subcategories as provided in the MIND dataset 
\\ [4mm]
\textbf{Calibration (complexity)} & Reading history & Continuous & article complexity \textbf{(1)} as calculated with the Flesch-Kincaid reading ease test 
\\ [4mm]
\textbf{Fragmentation} & Other users & Categorical & recommended news story chains \textbf{(2)}, which are identified following the procedure in {[}16{]} 
\\ [4mm]
\textbf{Activation} & Available articles & Continuous & affect scores, which is approximated by the absolute value of a sentiment analysis score \textbf{(3)}
\\ [4mm]
\textbf{Representation} & Available articles & Categorical & the presence of political actors \textbf{(4)}
\\ [4mm]
\textbf{Alternative Voices} & Available articles & Continuous & the presence of minority voices versus majority voices. We identify someone as a 'minority voice' when they are identified as a person through the NLP pipeline \textbf{(5)}, but cannot be linked to a Wikipedia page.\footnote{We acknowledge that this is a largely oversimplified approach towards identifying minority voices. This is a complex question that cannot be resolved to satisfaction within the scope of this paper.} \\
\bottomrule
\end{tabular}
\end{table*}

\paragraph{Representation} (Equation \ref{eq:rep})
aims to approximate a notion of viewpoint diversity (e.g. mentions of political topics or political parties), where the viewpoints are expressed categorically. Here $p$ refers to the presence of a particular viewpoint, and $P(p|S)$ is the distribution of these viewpoints within the overall pool of articles, while $Q^*(p|R)$ expresses the rank-aware distribution of viewpoints within the recommendation set. 

\paragraph{Alternative Voices} (Equation \ref{eq:alt})
is related to the Representation metric in the sense that it also aims to reflect an aspect of viewpoint diversity. Rather than focusing on the content of the viewpoint, it focuses on the viewpoint holder, and specifically whether they belong to a ``protected group'' or not. Examples of such protected/unprotected groups could be non-male/male, non-white/white, etc.\footnote{For more examples, see the UK 2010 Equality Act: \url{https://www.legislation.gov.uk/ukpga/2010/15/part/2/chapter/1}} This approach is based on the implementation of balanced neighbourhoods in recommender systems \cite{burke2018balanced}. With $m$ we refer to the distribution of protected vs.\ non-protected groups, with $m \in \{\mathit{Minority}, \mathit{Majority}\}$. ${P(m | S)}$ and ${Q^*(m | S)}$ refer to the distribution of these groups in the pool of available articles and rank-aware recommendation distribution respectively.

\smallskip\noindent%
Below is a summary of the formalization of DART with the RADio framework, the notation of which is defined in this section. In the next section, we show how to retrieve the necessary features from an example news dataset:

{\footnotesize
\begin{alignat}{3}
\mathit{Calibration} & = Cal\big(P^*(c | H), Q^*(c | R)\big) && = \sum_c Q^*(c | R) f \left( \frac{P^*(c | H)}{Q^*(c | R)} \right) \label{eq:cal} \\
\mathit{Fragmentation} & = Frag\big(P^*(e | R^{u}), Q^*(e | R^{v})\big) && = \sum_e Q^*(e | R^{v}) f \left( \frac{P^*(e | R^{u})}{Q^*(e | R^{v})} \right) \label{eq:frag} \\
\mathit{Activation} & = Act\big(P(k | S),Q^*(k | R)\big) && = \sum_k Q^*(k | R) f \left( \frac{P(k | S)}{Q^*(k | R)} \right) \label{eq:act}\\
\mathit{Representation} & = Rep\big(P(p | S), Q^*(p | R)\big) && = \sum_p Q^*(p | R) f \left( \frac{P(p | S)}{Q^*(p | R)} \right) \label{eq:rep}\\
\mathit{Alternative Voices} & = AltV\big(P(m | S),Q^*(m | R)\big) && = \sum_m Q^*(m | R) f \left( \frac{P(m | S)}{Q^*(m | R)} \right) \label{eq:alt}
\end{alignat}
}

\section{Experimental Setup} 
\label{sec:experiments}

In order to demonstrate RADio's potential effectiveness, we developed an NLP pipeline to retrieve input features to the metrics in Section~\ref{sec:metricsinpractice} and ran them on a public dataset. It should be noted that this pipeline is an imperfect approximation, and that each metric individually would benefit from more sophisticated methods.
The MIND dataset~\cite{wu2020mind} contains the interactions of 1 million randomly sampled and anonymized users with the news items on MSN News between October 12 and November 22 2019. Each interaction contains an impression log, listing which articles were presented to the user, which were clicked on and the user's reading history. The MIND dataset was published accompanied by a performance comparison on news recommender algorithms trained on this dataset,\footnote{Code available at \url{https://github.com/microsoft/recommenders/tree/main/examples/00_quick_start}} including news-specific neural recommendation methods NPA~\cite{wu2019npa}, NAML~\cite{wu2019naml}, LSTUR~\cite{an2019lstur} and NRMS~\cite{wu2019nrms}.  It was shown that these algorithms outperform general-purpose ones~\cite{wu2020mind} or common collaborative filtering models (such as alternating least squares (ALS)), in particular due to the short lifespan %
of news items~\cite{Garcin2013ALSNoGood}. These algorithms are trained on the impression logs in order to predict which items the users are most likely to click on. For the purpose of this paper we will evaluate these neural recommendation methods with the RADio framework (on the DART metrics) and compare their performance with two naive baseline methods, based on a reasonable set of candidates (the original impression log): a random selection, and a selection of the most popular items, where the popularity of the item is approximated by the number of recorded clicks in the dataset. 

\begin{table*}
\caption{Results for our RADio framework for recommendation algorithms on the MIND dataset. We use our preferred setup: JS divergence with rank-awareness @10. %
For interpretation of the results it should be noted that though a \emph{higher} score does imply higher divergence, this does not necessarily mean this is a \emph{better} score. Rather what it means to be better is dependent on the metric and the model chosen, for which we refer to Table \ref{tab:table_overview}.}
\label{tab:NDNRTable}
\resizebox{\textwidth}{!}{%
{\fontsize{7.4}{7.4} %
\begin{tabular}{lccccccc}
\toprule
\bf Algorithm & \textbf{\begin{tabular}[c]{@{}c@{}}Calibration\\ (topic)\end{tabular}} & \textbf{\begin{tabular}[c]{@{}c@{}}Calibration\\ (complexity)\end{tabular}} & \textbf{Fragmentation} & \textbf{Activation} & \textbf{Representation} & \textbf{Alternative voices} & \bf NDCG \\
\midrule
   LSTUR &              0.5847 &                   0.3632 &          0.9046 &   0.1819 &           0.1261 &                0.0409  & 0.4134 \\
    NAML &              0.5709 &                   0.3593 &          0.8836 &   0.1842 &           0.1230 &                0.0384 & 0.4091\\
     NPA &              0.5838 &                   0.3619 &          0.8979 &   0.1841 &           0.1359 &                0.0390 & 0.4068 \\
    NRMS &              0.5662 &                   0.3548 &          0.8872 &   0.1794 &           0.1278 &                0.0362 & 0.4163 \\
Most popular &              0.6526 &                   0.3477 &          0.8923 &   0.1949 &           0.1268 &                0.0342 & 0.2750 \\
  Random &              0.6636 &                   0.3981 &          0.9439 &   0.2715 &           0.2578 &                0.0698 & 0.2949\\
\bottomrule
\end{tabular}
}%
}
\end{table*}
\footnotetext{We computed NDCG for popular and random, and report on the original NDCG of the MIND publication for the neural recommenders, as it is more informative to the reader. We obtained similar results, since we only computed one inference cycle.}

Since RADio computes the average of all $\{P, Q\}$ pairs, we retrieve confidence intervals over  paired distances too, as illustrated in the sensitivity analyses below. In a traditional model evaluation setting, it would be desirable to generate confidence intervals via different model seeds or cross-validation splits. We refrain from doing this for our metric evaluation as this would introduce a multidimensional confidence interval (e.g., over $\{P, Q\}$ pairs and over model seeds).

We scrape articles via the URLs provided in the MIND dataset. Each article's metadata is enriched with five methods:
\label{sec:data_preparation}
\begin{enumerate}[leftmargin=*,nosep]
    \item \textbf{Complexity analysis} -- Each item is assigned a complexity score based on the Flesch-Kincaid reading ease test~\cite{kincaid1975derivation}, implemented in the Python module py-readability-metrics~\cite{readability}. Complexity is then discretized into bins, to accommodate for the discrete form of $D^*_f$.
    \item \textbf{Story clustering} -- The individual news items are clustered into so-called news story chains, which means that stories about the same event will be grouped together. This way, we add a level of analysis between individual news items and higher level categories (see Section~\ref{sec:metricsinpractice}). We use a TF-IDF based unsupervised clustering algorithm based on cosine similarity and a three days moving window, following the setup of \citet{trilling2020}.
    \item \textbf{Sentiment analysis} -- Using the textBlob open source NLP library we assign each article a sentiment polarity score~\cite{loria2018textblob}. Our focus is on the relative neutrality of articles, we thus take the absolute value of the negative / positive polarity score. %
    \item \textbf{Named entity recognition} -- Using spaCy, we identify the people, organizations and locations mentioned in the text \cite{spacy}, and count their frequency.
    \item \textbf{Named entity augmentation} -- For the entities identified in the text in the previous step, we attempt to link them to their Wikidata\footnote{\url{https://www.wikidata.org/}} entry through fuzzy name matching, to figure out if they are politicians, or in the case of organizations, political parties.\footnote{In the future one could also use additional data available on Wikidata for further refinement of the metrics, such as gender or place of birth / ethnicity for persons, industry type for organizations or country code for locations.}
\end{enumerate}

\noindent%
We implement RADio with the pipeline above. Table \ref{tab:table_implementation} links the numbered list above with the DART metrics. It provides an overview of the different metrics and their respective context distribution $P$ over normative concepts. The code for this implementation is available online.\footnote{\url{https://github.com/svrijenhoek/RADio/}}   %

We evaluate the outcome of our RADio framework for different recommender strategies (LSTUR, NAML, NPA, NRMS, most popular and random), with both KL Divergence and Jensen-Shannon as divergence metrics, with and without discounting for the position in the recommendation and at different ranking cutoffs.

\section{Results} 
\label{sec:results}
Having described our methodology and experimental setup around the operationalization of DART metrics, we analyze the results of the experiments on MIND. We separate descriptive analysis of the results in Section~\ref{sec:results} from the interpretation of normative interpretation of the metrics in Section~\ref{sec:discussion}. We choose to implement RADio with rank-awareness and JS divergence with a rank cutoff @N (the entire ranking list) as our default.
After commenting on the overall results, we further motivate that choice with a sensitivity analysis to different hyperparameters. We alter the divergence metric (KL or JS), rank-awareness (with and without a discount) and ranking cutoffs (@n, with $n = {1, 2, 5, 10, 20, N}$) for the different recommender models.

\begin{figure*}[t]
\includegraphics[width=0.8\textwidth]{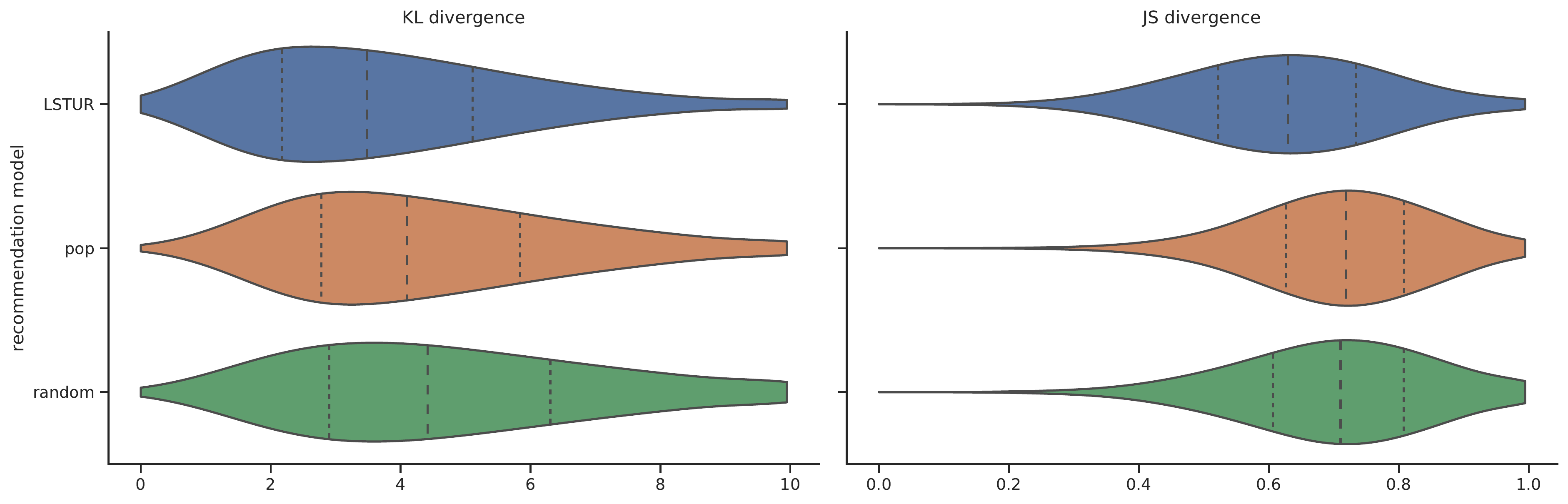}
\caption{Violin kernel density functions~\cite{ggplot2} over each $P, Q$ pair, for the Calibration (topic) rank-aware metric, rank cutoff @N (no cuttof), using KL (left) and the JS divergence (right). Thick and thin dashes show median and Inter Quartile Range (IQR) respectively.}
\label{fig:distance}
\end{figure*}

\begin{figure*}[t]
\includegraphics[width=0.8\textwidth]{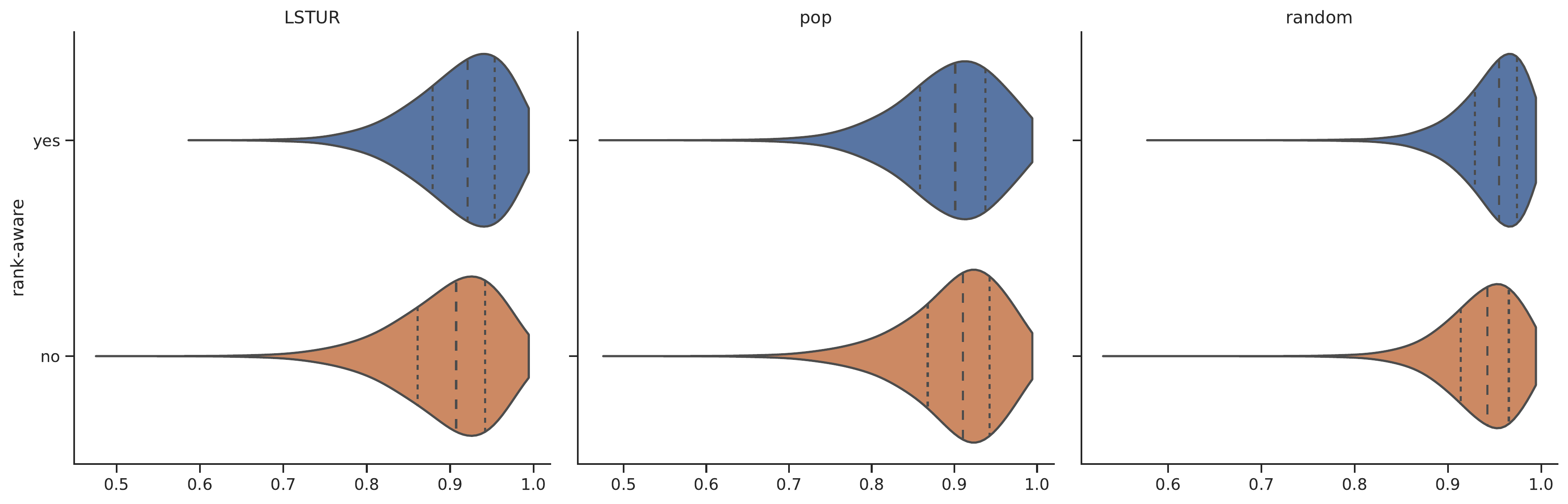}
\caption{Violin kernel density functions~\cite{ggplot2} over each $P, Q$ pair, for the Fragmentation metric with JS divergence and rank cutoff @N (no cuttof), on three different recommender approaches with (blue) and without (orange) rank-awareness. Thick and thin dashes show median and IQR respectively.}
\label{fig:discount}
\end{figure*}

Table \ref{tab:NDNRTable} displays results for RADio with rank-aware JS divergence.\footnote{More visualizations are available on the online repository} %
Higher values imply higher divergence scores, but whether high or low divergence is desired depends on the goal of the recommender system, which we will further elaborate in Section~\ref{sec:discussion}. %
The random recommender scores highest on divergence for all metrics and is also one of the least relevant by definition (see NDCG score). \textit{Most popular} and \textit{random} have comparable NDCG results. Popularity scores for the articles are derived from the clicks recorded in the MIND interaction logs, and many articles have zero or only one click recorded. %
When the candidate list contains exclusively articles with a similar number of clicks this forces the \textit{most popular} recommender to a random choice, which explains the artificial similarity between \textit{most popular} and \textit{random} in terms of the NDCG score.
Between the neural recommenders, most scores for LSTUR, NPA, NRMS and NAML are in lower ranges. Note that they produce similar recommendations (see NDCG values and \citet{wu2020mind}). Some notable differences can be observed when comparing these neural methods to the baselines. For example, we see that the neural recommenders are more Calibrated to the items present in people's reading history, though the most popular baseline performs marginally better in terms of Calibration of complexity. In the following, we further analyse the entire distribution of individual recommendation list divergences and test the sensitivity of RADio to different settings. Boxplots for all metrics and all recommender strategies are available in the online repository, where we highlight the importance of rank-awareness.

\subsection{Sensitivity to the Divergence Metric} 
JS divergence is our preferred implementation of universal diversity metrics. It is a proper distance metric and bounded between 0 and 1 (see Section~\ref{sec:metrics_divergence}). Figure \ref{fig:distance} substantiates that claim empirically, visualizing the sensitivity of RADio to the two described f-Divergence metrics: KL and JS Divergence. Clear differences can be observed in the distributions; KL divergence is skewed towards lower divergence, while JS divergence yields a more centered distribution of values. Additionally, JS divergence applies more contrast between the neural recommender systems and the naive recommendation methods and especially the random baseline. Due to the large sample in MIND, the random baseline is an approximation of a diverse recommendation set, given the candidate articles. In certain cases KL introduces consequential skew in the distribution of individual $P$, $Q$ comparison pairs across recommendation models; this does not occur to that extent with JS. %
Although KL Divergence is a well-known metric that can be found in many applications and is simpler to express mathematically, we found the JS divergence to be a better fit both theoretically and empirically. 

\subsection{Sensitivity to Rank-awareness}
In the original formulation of DART metrics \cite{vrijenhoek2021recommenders}, rank-awareness was not considered for most of the defined metrics. In our formalization, rank-awareness is the default. In Figure \ref{fig:discount}, we visualize the effect of removing the rank-awareness (in blue) on Fragmentation and compare to the original rank-aware Fragmentation (in orange). Rank-awareness allows for better differentiation between methods: LSTUR and ``most popular'' seem to be similarly distributed without a rank discount. Introducing rank-awareness shifts LSTUR towards a larger divergence, whereas ``most popular'' remains largely the same.

\begin{figure*}[h]
\includegraphics[width=0.85\textwidth]{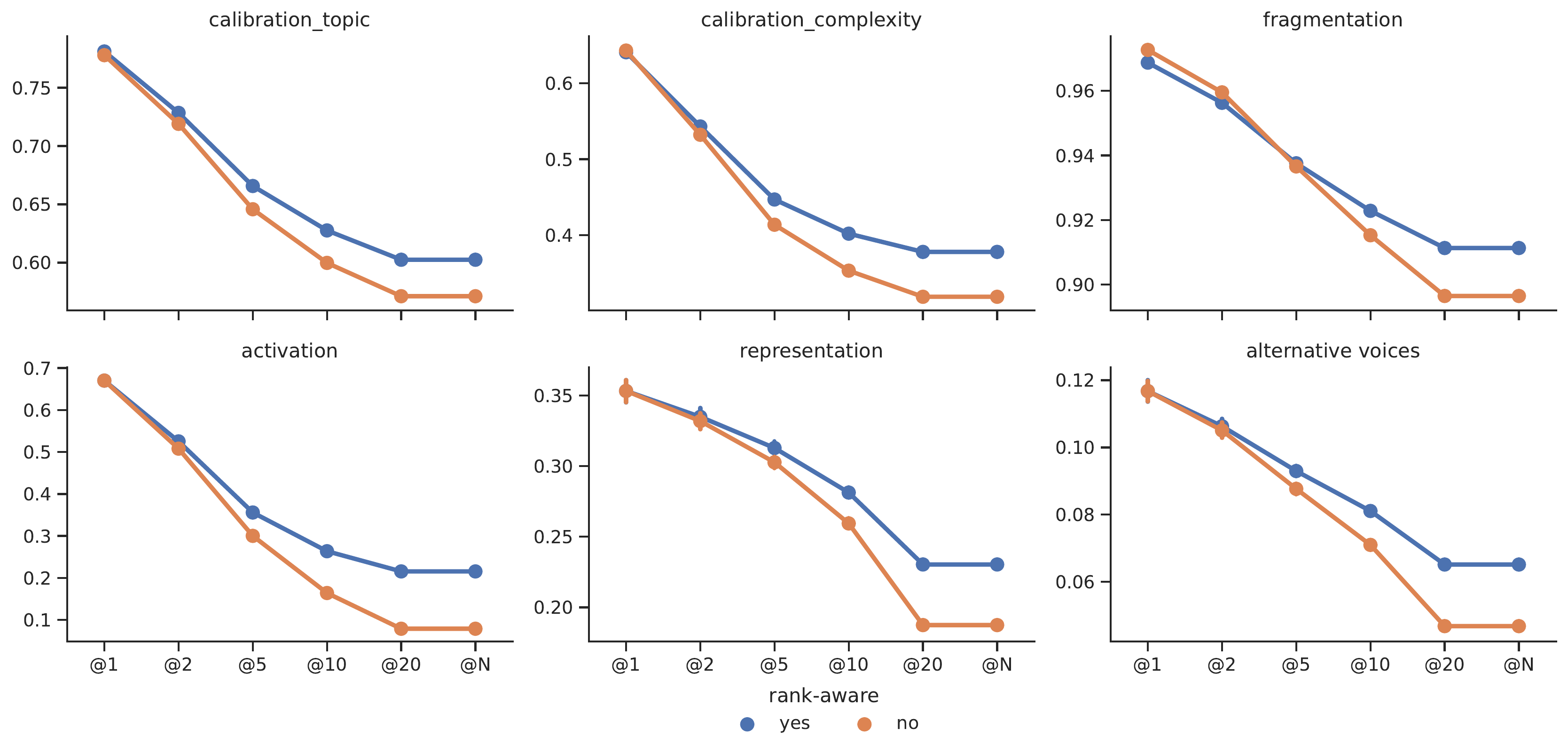}
\caption{Mean and 95\% confidence interval for each DART metric implemented with JS divergence for the LSTUR recommender. Sensitivity analysis of RADio on rank-awareness (blue and orange) and rank cutoff.}
\label{fig:rank-cutoff}
\end{figure*}

\subsection{Sensitivity @n}
One could also consider adding a cut-off point where only the top $n$ recommendations are considered for evaluation, the results of which are shown in Figure \ref{fig:rank-cutoff}. The figure shows that the effect of rank-awareness becomes stronger with a higher cut-off point, and causes the divergence score to stabilize after roughly 10 recommendations. This is because our MMR rank-awareness strongly discounts values further down the ranking.  @20 and @N (no cutoff) are similar for all metrics because MIND rarely contains more than 20 recommendation candidates. Note that when calculating the divergence score for Activation, Representation or Alternative Voices without rank-awareness and without cutoff point, there is no divergence to be reported as recommendation and target distribution are identical in these cases.\footnote{\url{https://github.com/svrijenhoek/RADio}}

\subsection{Normative Evaluation}
By comparing divergence scores across different recommender strategies, we can draw conclusions on the way they influence exposure of news to users. This is especially the case when comparing neural methods to the random recommender, which should reflect the characteristics of the overall pool of data. Combining this with DART's different theoretical models of democracy (summarized in Table \ref{tab:table_overview}), one can make informed decisions on which recommender system is better suited to one's normative stance than others. Imagine, for example, a public service media organization that aims to reflect Participatory norms and values in their recommendations. The Participatory model prescribes low Fragmentation and low Activation, which is shown in the scores of the neural recommenders. This would indicate that those models are more suitable for this organization's goals. In comparison, imagine a large media organization that wants to dedicate a small section of their website to Critical principles, consisting of one element with recommendations called ``A different perspective.'' The Critical model calls for a high divergence score in both Representation and Alternative Voices. Given that the random recommender scores best according to these principles, the neural recommenders would not be very suitable for this goal. Nevertheless, the conclusion that a random recommender is suitable for Critical norms and values is moot. Additional steps should be taken to further improve upon these scores: recommendation algorithm developers could tweak the trade-off between different target values in the recommendation, or even explicitly optimize on these metrics.

\section{Discussion}
\label{sec:discussion}
Choosing an f-Divergence score as the base for our metrics allows us to construct a single base formalization with a clear interpretation amongst all metrics; when the value is 0, the distribution between the recommendations and the chosen context is identical. The larger the measurements, the larger the divergence is. However, it also comes with a number of limitations. For one, f-Divergence does not take the relations between categorical values into account, and the ordering of the categorical values in the input vector is arbitrary. For example, two left-wing political parties in the Representation metric may be more similar than an extremely left-wing and an extremely right-wing party, but this is currently not taken into account. %
Related to this, in order to make continuous values suitable for our general discrete definition of f-Divergence, they need to be discretized into arbitrarily defined bins. This means that two very similar values may end up in different bins%
Future work may propose a different approach for calculating divergence between continuous variables.
Regarding the data enrichment pipeline, we identify a number of enhancement points.
While some metrics, such as topic Calibration, work with simple data on news topics that is often directly available in a dataset, other metrics require a more sophisticated data enrichment pipeline. The differences in these approaches appear in the results: the metrics with more trivial metadata retrieval setups show clear and distinct patterns for different recommender algorithms, but this is not the case for the more complicated ones. Furthermore, it is not possible to determine the quality of the pipeline, as we do not have a ground truth for evaluation. For future work, we suggest to take the base formalizations as constructed in this paper as a starting point, and work to improve the extraction of the relevant parameters for metrics such as Representation, Alternative Voices and Activation. Especially for the first two, there is already a large body of work that can facilitate this process~\cite{baden2017conceptualizing,draws2021assessing}. Human evaluation, including the input from editorial teams, would then be a promising way to evaluate these three normative metrics, similar to the work in the context of language generation bias~\cite{Dhamala2021fairMetrics}. Additionally, more insight needs to be gained on the influence of the choice of dataset. The MIND dataset contains a significant amount of so-called soft news, including articles on lifestyle, sport and entertainment, whereas the DART metrics are mostly applicable to hard news. %
The influence of the chosen dataset needs to be investigated in more detail, which can then lead to more informed decision-making on the trade-off between diversity and click-through rate, and what can reasonably be expected of a news recommender system.

\section{Conclusion}
\label{sec:conclusions}
In this paper we have made a first attempt at constructing and implementing new evaluation criteria for news recommender systems, with a foundation in normative theory. Based on the DART metrics, first theoretically conceptualized in earlier work, we propose to look at diversity as a divergence score, observing differences between the issued recommendations and a metric-specific target distribution. We proposed RADio, a unified rank-aware f-Divergence metric framework that is mathematically grounded and that fits several possible use cases within the original DART metrics and we hope beyond in future work. We showed that JS divergence was preferred over other divergence metrics. At first mathematically, as JS is a proper distance metric, and empirically, via a sensitivity analysis to different cutoff, rank-awareness and divergence metric regimes.
When our approach is adopted in practice, it enables the evaluation of news recommender systems on normative principles beyond user relevance. %
Finally, we wish to emphasize that the metrics proposed 
are meant to supplement standard recommender system evaluation metrics, in the same way that current beyond-accuracy metrics do. Most importantly, they are meant to bridge the gap between different disciplines involved in the process of news recommendation and to support more informed discussion between them. We hope for future research to foster interdisciplinary teams, leveraging each fields' unique skills and specialties.

\begin{acks}
We thank Max van Drunen, Natali Helberger, Maartje ter Hoeve, Dan Li and Marijn Sax for proof-reading. We also thank Ilias Koutsakis for helping cleaning up the code.
This research was partially funded by Bertelsmann SE \& Co. KGaA; by the Hybrid Intelligence Center, a 10-year program funded by the Dutch Ministry of Education, Culture and Science through the Netherlands Organisation for Scientific Research (\url{https://hybrid-intelligence-centre.nl}); the SIDN Fonds (\url{https://www.sidnfonds.nl/projecten/algorithms-for-freedom-of-expression-and-a-well-informed-public}).
All content represents the opinion of the authors, which is not necessarily shared or endorsed by their respective employers and/or sponsors.
\end{acks}

\bibliographystyle{ACM-Reference-Format}
\bibliography{main}

\end{document}